\documentstyle[12pt]{article}
\begin{document}
preprint: THU-98/03

\vskip 1.5 truecm
\centerline{\LARGE\bf{On the Gravitational Back Reaction
 }}\vskip .3 truecm
\centerline{\LARGE\bf{to Hawking Radiation.}}
\vskip 1.5 truecm
\centerline{{\bf S. Massar}}
\centerline{Institute for Theoretical Physics,}
\centerline{  Princetonplein 5,
P. O. Box 80 006, 3508 TA Utrecht, The Netherlands,}
\centerline{e-mail: S.Massar@fys.ruu.nl}
\vskip 1.5 truecm
\centerline{{\bf R. Parentani}}
\centerline{Laboratoire de Math\'ematiques et Physique Th\'eorique,
CNRS UPRES A 6083,}
\centerline{Facult\' e des Sciences, 
Universit\'e de Tours, 37200 Tours, France. }
\centerline{e-mail: parenta@celfi.phys.univ-tours.fr}
\vskip 1.5 truecm
\vskip 1.5 truecm

\centerline{\bf{Abstract}}

We show that a surface term should be added to the Einstein-Hilbert action 
in order to properly describe quantum transitions occurring around a black hole. 
The introduction of this boundary term has been advocated by Teitelboim and
collaborators and it has been 
used in the computation of the black hole entropy.
Here, we use it to compute the gravitational corrections to the transition
amplitudes giving rise to Hawking radiation. 
This
 surface term implies that 
the probability 
to emit a particle is
given by $e^{- \Delta A/4}$ where $\Delta A$ is the change in the area
of the black hole horizon induced by the 
emission.
Its inclusion 
at the level of the amplitudes 
therefore
 relates 
quantum black hole radiation to the 
first law of black hole
dynamics. In both cases indeed, the term 
expressing the change in area
directly results from 
the same boundary term introduced for the
same reason: to obtain a well defined action principle.
\newpage
\section{Introduction}

There are two possible approaches to the 
gravitational back reaction to Hawking radiation. The first is to
develop a microscopic theory of quantum gravity and to use it to
calculate the properties of black holes. This program has been
partially realized in the context of super-string theory\cite{calmad}. 
The second is to use Hawking's calculation\cite{H} as a starting point to compute
gravitational corrections to black hole evaporation.
 Hopefully these two
approaches  should meet in some middle ground.

In the more conservative second approach, the back reaction has been
addressed along two complementary lines. The first is the
``semi-classical'' theory wherein the metric remains classical, but
one takes as source in Einstein's equations the mean 
(average) 
energy-momentum
of the second quantized fields propagating on this self consistent
background\cite{Bard}\cite{massar}. 
This describes the mean back reaction to Hawking radiation
and 
confirms 
that black holes evaporate 
adiabatically
in a time $O(M^3)$ as first
predicted by Hawking\cite{H}.

The second line of attack is to take into account the dynamics
of gravity at the level of transition amplitudes before performing
any quantum averaging. 
In the approach initiated by Keski-Vakkuri, 
Kraus and Wilczek (KKW) \cite{KVKW}, one 
replaces the matter action $S_{matter}$ in a given geometry 
by the self consistent action of matter plus gravity
$S_{matter+gravity}$. 
One then postulates that the wave functions 
governing transition amplitudes are WKB, i.e. are of the
form $e^{i S_{m+g}}$. 
This approach closely corresponds to the
computation of the transition amplitudes in quantum cosmology
 performed in \cite{wdwpt}. Indeed,
the gravitational waves appearing in matrix elements 
are WKB solutions of the Wheeler-De Witt equation
and thus have also their phases determined by
the matter energy of the quanta involved in the transition.
Another useful analogy to this approach 
is provided by the quantum description of the
trajectory of an accelerated detector experiencing the Unruh effect\cite{U}.
In that case, the given trajectory (which plays the role of  the 
classical geometry in black hole physics)
has been replaced by WKB waves
so as to take into account recoil effects according to Feynman rules\cite{rec}\cite{suh}.

The aim of the present work is to 
clarify the mechanisms at work in the approach of   KKW.
In particular, we determine
what is the correct gravitational action that must be used
upon describing black hole radiance.
There is indeed some ambiguity in the choice of the boundary terms
at the horizon. 
On physical grounds, we select the appropriate one. This amounts to work
with the boost parameter defined at the horizon 
as the time parameter 
instead of the time defined at infinity.
In this, our treatment of the boundary terms closely follows the 
work of Ba{\~ {n}}ados, Teitelboim and Zanelli\cite{BTZ}.
 
In a former paper\cite{MP}, we already used their improved action
in a Euclidean context. Our aim was to compute
self consistently the action of instantons 
governing probability transitions in the presence of
horizons. The main advantage of the Euclidean approach
lies in its simplicity. In that case indeed, the choice of the 
correct action is particularly transparent. 
However
this approach has an important weakness:
one must postulate, as in all 
works 
based on the
Euclidean path integral see e.g. \cite{HHR}, 
that no conical singularity be present in the Euclidean geometry. 
This condition should be proven from first (quantum) principles and not assumed. 
Remember that in the absence of gravitational
back-reaction it is the fact that the quantum state is vacuum that 
implies Green functions possessing well defined Euclidean
properties.

The present work, that of KKW, and that of \cite{MP} all reach the
same conclusion.
By using the Einstein-Hilbert action rather than only the matter
action as the phase of WKB waves, one finds that the rate of emission of
particles of energy $\lambda$ from a black hole of mass $M$ is
\begin{equation}
R_{M \to M - \lambda} = N(\lambda, M ) e^{- \Delta A (\lambda, M) /4}
\label{R}\end{equation}
where $\Delta A (\lambda,M)= A(M) - A(M-\lambda)$ 
is the change of area of the black hole due
to the emission of a particle of energy $\lambda$, and $N$ is a phase
space (also called grey body) factor which cannot be calculated 
in this approximation scheme. 
To first order in $\lambda$,
$\Delta A (\lambda, M) /4 = \lambda / 8 \pi M$ and one recovers
Hawking's canonical result characterized by a temperature.

One of the main interests of eq. (\ref{R}) is to 
suggest
that the thermal properties of Hawking radiation 
and the first law of black hole mechanics  both stem from the same 
fundamental principle
valid at the quantum level, i.e. before averaging. This would explain
why they give a consistent description of the thermal properties of
black holes.
We recall that the first law relates neighboring black hole
classical configurations and as such seems to be completely
independent of
quantum black hole radiance. 
A first direct indication of the relationship 
between the first law and black hole evaporation 
is provided
by the instantonic  approach\cite{MP}. 
Indeed in this work,  $\Delta A (\lambda, 
M) /4$ has exactly the same origin as the $A/4$ term 
in the partition function of a black hole calculated using
Euclidean path integral techniques\cite{GH}\cite{BTZ}. In 
both these cases, this term arises from a boundary term which is
introduced at the horizon in order to 
select the Hamiltonian variational
principle appropriate for the physical problem under investigation. 
And in the first law as formulated by Wald \cite{W}, 
$dA/4$ is also equal to the variation of the same boundary term at the
horizon.
At this point it should be noticed that 
this boundary term does not appear in the calculation of KKW. 
Our aim is therefore to carefully reformulate the calculation 
of transition amplitudes 
to establish that the boundary term must be present right from the beginning.

We finally note that the present procedure applies to 
all processes occurring in the 
presence of horizons and leads to transition rates 
determined by the corresponding changes in area. 
In particular it also applies to the transitions of accelerated systems. 
This suggests that there might exist a quantum statistical extension
of the thermodynamical approach to general relativity\cite{jac}
which is valid everywhere
and not only applicable to black hole horizons.
In this case one would obtain a kind of universal 
``holographic principle''\cite{HOL} valid 
for all horizons and expressing the changes of the
effective degrees of freedom to which systems
living on one side are coupled.

The plan of the paper is as follows:
we  begin by reviewing the standard derivations of Hawking radiation
without back reaction, following \cite{H} and \cite{U}. 
We then review the role of boundary terms in
the Einstein-Hilbert action. Finally we rederive eq. (\ref{R})
in two 
complementary ways. The first
emphasizes the role of 
detectors which measure the presence
of particles and which deform differently the geometry according to their
quantum state. The second way only makes use of the regular
structure of the wave functions near the horizon when expressed
in advanced Eddington-Finkelstein coordinates. 

\section{Hawking Radiation}\label{two}

In this section we fix the notations and review two standard
derivations of black hole radiation. 
In the first, following Unruh's work\cite{U}, one introduces
a particle detector whose position is fixed and one
determines the populations of quanta from its
transition rates.
In this way, one only uses basic quantum mechanical rules.
The second approach is more intrinsic 
and closer to the original Hawking's 
derivation\cite{H}. Black hole radiation is established 
through the Bogoljubov transformation relating in-modes which
determine the Heisenberg state and out-modes which
define the on-shell particles found at infinity.
The specification of the in-modes will be obtained
by imposing well defined analytical properties on the
horizon\cite{U}\cite{DamourRuff}.

The metric of a Schwarzschild black hole is
powerful\begin{equation}
ds^2 = -(1 - {2 M \over r}) dt^2  +(1 - {2 M \over r})^{-1} dr^2 + r^2
d\Omega^2
\end{equation}
We introduce the light like coordinates $u$ and $v$:
\begin{equation}
v,u = t \pm r^*\quad , \quad r^* = r +  2M \vert\ln (r- 2M) \vert
\end{equation}
and the Kruskal coordinates $U$ and $V$:
\begin{equation}
U = - 
{ 1 \over \kappa} e^{-\kappa u} \quad , \quad 
V= { 1 \over \kappa} e^{\kappa v} 
\end{equation}
where $\kappa= 1/4M$ is the surface gravity.

For a black hole formed by the collapse of a star, the outgoing modes
solution of the Klein Gordon equation
have the following form near the horizon:
\begin{equation}
\phi_{\omega, l, m} = { e^{-i\omega U} \over \sqrt{4 \pi \omega}}
{Y_{lm}(\Omega) \over r} \label{Umodes}
\end{equation}
Further from the horizon this is not an exact solution because of the
potential barrier which surrounds the black hole. For simplicity in
this article we shall neglect the potential barrier. It would only
encumber the expressions, whereas all the physics lies near the
horizon where it 
plays no role.

A  way to understand the structure of the modes
eq. (\ref{Umodes}) is that they correspond to the absence of particles
as seen by an infalling observer. 
Indeed the proper time lapses of an
infalling observer are proportional to 
$\Delta U$.
This is directly apparent by 
re-expressing the Schwarzschild metric in $U, V$ coordinates.
For $r- 2M\ll 2M$, one finds $ds^2 \simeq - dUdV + r^2 d^2 \Omega$
which explicitizes the initial character of $U$  near  $r=2M$. 

The field operator
can 
be decomposed as:
\begin{equation} \Phi = \sum_{\omega,l,m} a_{\omega,l,m}
\phi_{\omega, l, m}  + {\rm h.c.} + {\rm ingoing\ modes} 
\end{equation}
By definition, 
the Unruh vacuum state $|0_U>$ is annihilated by the $a_{\omega,l,m}$
operators.

Consider now a particle detector at fixed radius $r$ and angular
position $\Omega$ which has two levels $|e>$ and $|g>$ of
energy $E_e$ and $E_g$ with $\Delta E = E_e - E_g >0$. 
Its coupling to the field $\Phi$ is given in
interaction representation by
\begin{equation}
H_{int} = \gamma
\Phi(t,r,\Omega)\left( e^{+i \Delta E t} |e><g| + {\rm h.c.} \right)
\end{equation} 
where $\gamma$ is a coupling constant.
Initially  the detector is in its ground state, hence the combined
state of the detector and field is $|0_U>|g>$. 
In the interacting picture, at late times and to first
order in $\gamma$, the state of the detector plus field is:
\begin{eqnarray}
|\psi_g>&=&|0_U>|g> - i \gamma \int dt e^{i \Delta E t} \Phi(r,\Omega,t)
|0_U>|e>\nonumber\\
&=&|0_U>|g> - i \gamma \sum_{\omega l m}\int dt e^{i \Delta E t} e^{-i
\omega  C e^{-\kappa t} } {Y_{lm}(\Omega)\over r \sqrt{4 \pi \omega}}
a^\dagger_{\omega l m}
|0_U>|e> \label{gtoe}
\end{eqnarray}
where $C= (1/\kappa) \exp (\kappa r + {1 \over 2}\ln (r - 2M))$.
Similarly if the detector is initially in its excited state, the
amplitude of the detector plus field at late times is
\begin{equation}
|\psi_e>=|0_U>|e> - i \gamma \sum_{\omega l m}\int dt e^{-i \Delta E t} e^{-i
\omega C e^{-\kappa t}}  {Y_{lm}(\Omega)\over r \sqrt{4 \pi \omega}}
a^\dagger_{\omega l m}
|0_U>|g>\label{etog}
\end{equation}
The ratio of the amplitude of getting excited 
to the amplitude of getting
deexcited is
\begin{equation}
{{\cal{A}}_{g \to e, \omega,l,m} \over {\cal{A}}_{e \to g, \omega,l,m}}
={ \int dt e^{+i \Delta E t} e^{-i
\omega  C e^{-\kappa t}} \over \int dt e^{-i \Delta E t} e^{-i
\omega C e^{-\kappa t}}}\end{equation}

By replacing $t$ by $t + i \pi/ \kappa $ in the upper integral,
one obtains
that $A_{g \to e} = A_{e \to g}^* e^{- \pi \Delta E /\kappa}$, see \cite{PB}\cite{suh}.
Thus the 
ratio of the probabilities of transition is:
\begin{equation}
{|{\cal{A}}_{g \to e, \omega,l,m}|^2 \over |{\cal{A}}_{e \to g, \omega,l,m}|^2}
=e^{- {2\pi \over \kappa} \Delta E} \label{Prob}
\end{equation}
corresponding to the rates in a thermal bath at temperature $
\kappa / 2 \pi = 1/8 \pi
M$. In equilibrium the ratio of rates is equal to the ratio of
probabilities to be in the excited or ground state.
This derivation is equivalent to calculating the 
Bogoljubov transformation between Unruh modes and Schwarzschild
modes. Moreover, it provides a physical interpretation to individual 
Bogoljubov coefficients as transition amplitudes\cite{U}\cite{PB}. 

In the second way,
the spectrum of  emitted particles is
determined 
in terms of modes which are  eigenvectors of $i\partial_t=\lambda$. There are
two modes for each value of $\lambda$ which are non vanishing either
outside the horizon or inside the horizon:
\begin{eqnarray}
\varphi_{\lambda,l,m,\pm} &=& {e^{- i\lambda u}\over \sqrt{4 \pi \lambda}} 
{ Y_{lm}(\Omega) \over r}
 \theta (\pm(r-2M)) 
\end{eqnarray}
Where we have once more neglected  the
potential barrier outside the black hole.
The Schwarzschild modes $\varphi_+$ 
are those which would naturally be used by a static
observer around the black hole to describe the presence or absence of
particles. 
On the other hand the modes $\phi_{\lambda,l,m}$ associated to the Unruh vacuum
state are linear combinations of $\varphi_+$ and $\varphi_-$. To
determine this linear combination one must impose that the
$\phi_{\lambda,l,m}$ have only positive frequency for an infalling
observer.
One way to construct the $\phi_\lambda$ modes is to express the
Schwarzschild modes in Kruskal coordinates:
$\varphi_{\lambda \pm} \simeq (\mp U)^{i \lambda / \kappa}$. Since 
$\Delta U$ is
proportional to the proper time of an infalling observer,
$\phi_\lambda$ must be the linear combination of $\varphi_{\lambda
\pm}$ which is analytic and bounded in the 
lower half of the complex
$U$ plane:
\begin{equation}
\phi_{\lambda,l,m} = {1 \over \sqrt{ 1 - e^{- 2 \pi \lambda/ \kappa}}}
\left( \varphi_{\lambda,l,m,+} + e^{-  \pi \lambda /\kappa}
\varphi_{\lambda,l,m,-} 
\right )\label{US}\end{equation}
where the normalization factor ensures that the $\phi$ modes have unit
norm.
The 
weights in
eq. (\ref{US}) define  the Bogoljubov coefficients $\alpha_\lambda$
and $\beta_\lambda$.
Their ratio satisfies
\begin{equation}
{ | \beta_ \lambda |^2 \over | \alpha_ \lambda |^2}
= e^{- 2 \pi  \lambda/\kappa }
\end{equation}
This implies that the Unruh
vacuum is a thermal distribution of Schwarzschild particles 
at temperature $\kappa /2  \pi M$ (compare with eq. (\ref{Prob})).

We will find it convenient below to use the same argument, but
rephrased in
 Eddington-Finkelstein coordinates
$v,r,\Omega$ in which the metric has the form:
\begin{equation}
ds^2 = -(1-2M/r) dv^2 
+ 2dvdr + r^2 d\Omega^2
\end{equation}
Near the horizon the metric takes the simple form
$ds^2 \simeq 2dvdr + r^2 d\Omega^2$ which shows that $v,r$ are
inertial coordinates near the horizon. 
An infalling observer follows, near the horizon, approximately the
trajectory $v=const$, so that $-r$ is approximately a proper time
parameter along his world line.
$\phi_{\lambda,l,m}$ must have only positive frequency in $-r$ at fixed
$v$.
Near the horizon, in Eddington-Finkelstein coordinates, the
$\varphi$ modes have 
the form  $\varphi_{\lambda,l,m,\pm} \simeq
e^{- i{\lambda \over \kappa }(\kappa v - \ln (r-2M))} \theta (\pm(r-2M))$. 
The regularity condition imposes
that to obtain a $\phi$ mode, one analytically continues $\varphi_+$ in
the upper half complex $r$ plane to obtain the relative weight of the 
$\varphi_+$ and $\varphi_-$ modes, as shown in \cite{DamourRuff}. 
In this way one recovers eq. (\ref{US}).

\section{Boundary terms in the Einstein-Hilbert Action}

In this section we examine the boundary terms in the Einstein-Hilbert
action. We start our analysis with the action expressed in Hamiltonian
form:
\begin{equation}
S = \int dt \left\{
\pi^{ab} \dot g_{ab} + p \dot q - N H - N^i H_i
\right\}
 \label{S}
\end{equation}
where $g_{ab}$ is the spatial metric, $\pi^{ab}$ its conjugate
momentum, $q$ and $p$ the coordinates and momentum of matter, $N$ and
$N^i$ the lapse and shift, and $H$ and $H_i$ the energy and momentum
constraints. 

We shall consider this action defined in the right  quadrant of the full
Kruskal manifold, and take the time foliation to resemble Schwarzschild
time. The equal time slices then have two boundaries: one at spatial
infinity and one at the horizon (more precisely at the intersection of
the past and future horizon). Variations of the action eq. (\ref{S})
gives rise to boundary terms because of the presence of spatial
derivatives of the metric in $H$ and $H_i$. In order that the action be
extremal on the solutions of the equations of motion, one must add
surface terms to $S$ that cancel the boundary terms. We now review
these boundary terms, starting with the boundary at infinity. Details
of the calculations will not be presented. They can be found in
in many papers, see for instance 
\cite{RT}\cite{K}\cite{BTZ}\cite{T}.

Varying  eq. (\ref{S}) while imposing that the space time is
asymptotically flat yields 
\begin{equation}
\delta S = \int dt {\rm\ terms\ giving\ eqt.\ of\ motion} + t_\infty \delta M_{ADM}
\end{equation}
where $t_\infty = \int dt N $ is the proper time at infinity
and $M_{ADM}$ is the mass at infinity (defined in terms of the
behavior of $g_{rr}$ for large $r$).
In this form the action is extremal on the equations of motion if one
varies among
the class of metrics for which $M_{ADM}$ is kept fixed. 

On the contrary,  if one varies among the class of metrics
for which $t_\infty$ is kept fixed but $M_{ADM}$ is arbitrary
one must 
subtract
 the surface term $M_{ADM} t_\infty$ to the action 
$S$ so that
the variation of the new action $S'$  yields
\begin{equation}
\delta S' = \int dt {\rm\ terms\ giving\ eqt.\ of\ motion} - 
M_{ADM} \delta t_\infty
\end{equation} 
This second form of the action is generally more convenient because  
one can calculate $S'$ as a
function of $t_\infty$, and then use this expression to calculate 
the time evolution of the matter and metric. 
Indeed since $M_{ADM}$ is a conserved quantity, $\partial S' /
\partial 
M_{ADM} = const$ gives the
time evolution.
A second more physical reason is that one may want to compare the
evolution of systems with different ADM masses, and hence one does not
want to fix it to start with.
However we must emphasize that there is nothing fundamentally wrong
with the variational principle based on $S$. This will be important
below.

We now turn to the boundary term at the horizon. The analysis proceeds
exactly in parallel with the proceeding one. The form of the boundary  term is
dictated by the fact one       requires
 that near the horizon the lapse and shift vanish (or equivalently
that at $\rho =0$, the momentum $\pi^\rho_\rho$ and the derivative of
the area of the surfaces of constant $\rho$, $\partial_\rho A$, 
vanish, see \cite{T})
and hence the metric can be put in the form 
\begin{equation}
ds^2 = -\rho^2 \kappa^2 dt^2 + d \rho^2 + r^2(\rho) d\Omega^2
\label{metric}
\end{equation}
For a Schwarzschild black hole $\rho = \sqrt{8 M (r - 2 M)}$, 
$\kappa = 1/4M$ and $t=t_{\infty}$.

Upon varying eq. (\ref{S}) one then finds a boundary term at the
horizon
\begin{equation}
\delta S = \int dt {\rm\ terms\ giving\ eqt.\ of\ motion} 
- 
\Theta \delta A/8 \pi
\end{equation}
where $\Theta = \int \kappa dt$ is the hyperbolic angle 
and $A$ is the area of the horizon.
The action in this form is therefore extremal on the equations of
motion for the class of metrics which have fixed  horizon area.

One can add  to this action a surface term at the horizon
and define
$S''=S 
+ \Theta A /8 \pi $ so that
its variation takes the form:
\begin{equation}
\delta S'' = \int dt {\rm\ terms\ giving\ eqt.\ of\ motion} +
A \delta \Theta / 8 \pi
\end{equation}
This action is extremal on the equations of motion provided one makes
variations
in the class of metrics which have fixed $\Theta$ at the horizon. 

This is the action that must be used when calculating the action of
the Euclidean continuation of the black hole\cite{BTZ}. 
Indeed regularity of the
Euclidean manifold at the horizon imposes that the Euclidean angle
$\Theta_E = 2 \pi$. The surface term in the action then contributes a
term $A/4$ to the partition function which is 
interpreted as the entropy of the
black hole. This is also the action that was used in \cite{MP}
to calculate the self-consistent actions of instantons.
In the present paper, we shall use it in the Lorentzian sector.
The new aspect brought in by its use
is that the time evolution will be given 
in terms of the boost parameter
$\Theta$ rather than the time at infinity.

Thus upon considering processes occurring around a black hole,
there are a priori 4 
actions that can
be considered according to choice of the surface terms.
Two however are unphysical. Indeed fixing both the ADM mass
and the horizon area is inconsistent. For instance for the vacuum
spherically symmetric solutions, fixing the ADM mass determines the horizon
area. Similarly fixing both the time at infinity and the hyperbolic
angle $\Theta$ is inconsistent. Thus one is left with two
possibilities: fixing the ADM mass and $\Theta$ or fixing $t_{\infty}$
and $A$. 
A more mathematical reason why these are the only 
two possibilities is that the constraints $H=0$ and $H_i=0$ 
viewed as differential equations  
need boundary conditions in order to yield a unique solution
and fixing the ADM mass or the horizon area but not both provides the required 
boundary data\cite{T}.

The choice among these two possibilities is dictated by physical
considerations. If the ADM mass is not fixed but $A$ is, 
this means that one is in
effect considering situations in which there are exchanges of energy
between the matter surrounding the black hole and infinity while
leaving the black hole itself unchanged. On
the other hand if one fixes $M_{ADM}$ while letting $A$ vary,
one is comparing situations in which the black hole and the
surrounding matter exchange energy, but no energy is exchanged with
infinity.

Clearly the process of black hole evaporation belongs to the second
situation. Therefore in the next section we shall consider the action
\begin{equation}
S = \int dt \left\{ \pi^{ab} \dot g_{ab} + p \dot q - N H - N^i H_i 
\right\}
+ \Theta A / 8 \pi
\label{S2}
\end{equation}
This action gives the time evolution of matter as a function of
$\Theta$ for different values of $A$ with $M_{ADM}$ fixed.

\section{Gravitational Back Reaction to Hawking Radiation}

In this section we calculate the new expressions
of the
different wave functions introduced in section \ref{two}
by replacing the matter action in the given Schwarzschild 
geometry by 
the gravitational action eq. (\ref{S2}).
We then to use these
wave functions to compute the gravitational corrections  to
black hole radiation.

We start with the description of black hole radiation based
on the readings of a static detector. It is through the change of the  
deformation of the geometry induced by the 
change of the quantum state of the detector that 
the change in area will appear in the transition rates.

We first compute the new time dependence of the 
wave functions associated with the two states of the 
detector. 
Since the detector is at $r=const$, the
geometry is static  and the $p \dot q$ and $\pi^{ab}\dot g_{ab}$
terms in the action vanish. 
Moreover, on-shell, the constraints also vanish.
Hence the
only term contributing to the total action (matter + gravity)
is the surface term at the
horizon. This term is equal to $\Theta A(g)/8 \pi $ or $\Theta A(e)/8 \pi$
where $A(g)$ ($A(e)$) is the horizon area when the detector is in the
ground (excited) state. 
Thus the time dependence of the wave functions 
are 
\begin{eqnarray}
\psi_{g}(\Theta) &=& e^{i \Theta
A(g)/8 \pi}\nonumber\\
\psi_{e}(\Theta) &=& e^{i \Theta
A(e)/8 \pi}
\end{eqnarray}

It remains to determine the new expression for the 
outgoing modes which replaces eq. (\ref{Umodes}). 
As in the absence of back-reaction, the structure of 
the modes 
must be such that they determine the vacuum near the 
horizon. Since $\Theta$ has been defined at the
horizon, its relationship to the inertial light like 
coordinates $U_{hor}, V_{hor}$ also defined at the horizon is of the 
form $U_{hor} = 
-e^{- \Theta}$. This encodes the exponential
Doppler shift which is the hall mark of horizons. 
Therefore, the new expression is 
\begin{eqnarray}
\phi_{\omega}(\Theta,r) = D
e^{iC \omega e^{-\Theta}}
\end{eqnarray}
in the place of $e^{iC\omega e^{-\kappa t}}$ see eqs. (\ref{gtoe}, \ref{etog}).

As in Section 2, 
the transition amplitudes are given by the 
``time'' integral of the product of the three waves.
Up to the {\it same} overall constant, they are given by
\begin{eqnarray}
{\cal{A}}_{g \to e + \omega} &=& 
\int d \Theta\ \psi_{g}(\Theta)
\psi_{e}(\Theta)^* \phi_{\omega}(\Theta)^*
\nonumber\\
{\cal{A}}_{e \to g + \omega} &=& 
\int d \Theta\ \psi_{g}(\Theta)^*
\psi_{e}(\Theta) \phi_{\omega}(\Theta)^*
\label{exp}\end{eqnarray}
Note that the area of the horizon decreases if the detector is in its
excited state. Therefore, in the limit of small $\Delta E$,
i.e. in the test particle limit,
the factor $\psi_{g}(\Theta)^*
\psi_{e}(\Theta) = e^{i \Theta (A(e) - A(g))
/8 \pi}$ tends to the background
field expression $e^{-it \Delta E}$
with the correct sign of the phase.

Using the new expressions eq. (\ref{exp}), by replacing
$\Theta$ by $\Theta + i \pi$ in the upper amplitude, one obtains
\begin{equation}
{ | {\cal{A}}_{g \to e + \omega}|^2 \over |{\cal{A}}_{e \to g + \omega}|^2 }
= e^{- (A(e) - A(g))/4}
\end{equation}
It is thus $(A(e) - A(g))/4$, 
the difference of the horizon
areas if the detector is in its excited or ground state
which governs the equilibrium distribution of the detector's states.
This describes a micro-canonical distribution since we are considering exchanges
of energy between the black hole and the detector with no mass exchange 
at infinity. This distribution replaces the canonical expression of 
eq. (\ref{Prob}),
governed by the energy change $E_e - E_g$ and 
Hawking's temperature $\kappa / 2 \pi$.

The new ratio of the transition rates is a function of the change in area 
because the integrand of the transition amplitude
contained the product $\psi_{g}^* \psi_{e}$ which 
describes the replacement of one classical solution
by the other one. 
It is indeed through this $\psi^* \psi$ product 
that the change in area entered into the expression since
its phase is given by the {\it difference} of the total matter+gravity actions.
As noted above this difference tends to $- \Delta E t$ when $\Delta E$
is small. This recovery of the background field wave functions in the
limit of small differences of matter energy is a generic
 feature\cite{bfa} of
taking into account a neglected degree of freedom (here gravity). The
same mechanism arises upon taking into account recoil effects 
 in scattering amplitudes,
see \cite{rec}. It also explains why
 one recovers the conventional
expressions of transition amplitudes starting from
the solutions of the Wheeler-DeWitt equation in quantum cosmology\cite{wdwpt}.

In this calculation the change in area is due to the change
of the 
quantum state of the detector. However the existence of a detector
is not intrinsic to black hole radiation: the detector was only used to reveal
the existence of the quanta.
Therefore we seek for an intrinsic derivation of 
black hole radiance in which the change in area 
is due to the emission process itself.
To this end, we must introduce a model for the matter waves
which takes into account the deformation of the gravitational 
background. The
simplest model is that of  KKW in which the matter is described as
a spherically symmetric (s-wave) light-like shell. 
The main result of \cite{KVKW}
is that outside and inside the shell the metric is Schwarzschild, but
with the mass parameter $M_{ADM}$ and $M_{ADM} - \lambda$ respectively
where
$\lambda$ is the energy of the shell.
Then the time parameters inside
and outside the shell are different and 
related by junction conditions on the shell.
In what follows this relationship will play no role. 
The shell follows a light like
geodesic in both the inside and outside metric. This property allows
us to calculate the action of the shell in a straightforward manner.

Inside the shell the metric is
\begin{equation}
ds^2 = - (1- 2M'/r) dt^2 + (1-2M'/r)^{-1} dr^2 + d\Omega^2 
\end{equation}
where $M'= M_{ADM}-\lambda$ is the 
final
 mass of the black hole.
The
shell follows a geodesic in this metric.  
\begin{eqnarray}
{d r_{sh} \over d t }&=&  (1- 2M'/r_{sh}) \nonumber\\
t_{sh} (r, M') &=&  r + 2 M'
\ln (r - 2 M') = r + {1 \over 2 \kappa'}  \ln (r - 2 M')
\end{eqnarray}
where $\kappa' = 1/ 4 M'$.

As in the previous treatment, it is particularly 
convenient to reexpress the evolution in terms of the hyperbolic
angle $\Theta$ using the Jacobian $d \Theta / dt = \kappa'$:
\begin{equation}
\Theta_{sh} (r, M') = \kappa' r + { 1 \over 2}
\ln (r - 2 M') \label{th}
\end{equation}
Notice that close to the horizon, the description of the 
shell's trajectory is now expressed only in terms of 
quantities locally defined, i.e. insensitive to the matter
distribution at larger radii.

The  action of the shell can in principle be calculated by solving
the Hamilton-Jacobi equations. Since $A'$ is a constant of motion,
these can be solved by separation of variables to yield the form
$S(A', \Theta) = A'\Theta / 8 \pi + f(r,A')$.
Furthermore $\partial S / \partial {A'}=0 $ 
must give the equations of motion which implies that
\begin{equation}
{\partial f \over \partial{A'} }
=  -  {\Theta_{sh} (r, M') \over 8 \pi}
\end{equation}
Inserting eq. (\ref{th}), this can be integrated to yield
\begin{equation}
f(A') =
 \int^{A}_{A'} 
{d \tilde A\over 8 \pi } ( {1 \over 2} \ln (r - r_{hor}(\tilde A)) + 
{\kappa(\tilde A) r } )
\end{equation}
where we have introduced $r_{hor}(A)$, the radius of the horizon 
parameterized by its area. Similarly we have introduced $\kappa( A)$.
Note however that $\kappa$ is not strictly speaking the surface
gravity of the black hole, rather it is determined by the form of the
metric near the horizon eq. (\ref{metric}).

In order to obtain the action which governs the 
{\it transition} from $A$ to $A'$, we must subtract from $S(A')$
the action $A \Theta / 8 \pi$ which describes the geometry in the absence of the
shell. This 
yields \begin{equation}
S_{shell}(A', A) = -  \Theta (A - A')/ 8 \pi +  
\int^{A}_{A'} {d \tilde A\over 16 \pi } \ln (r - r_{hor}(\tilde A)) +...
\label{Sl}
\end{equation} 
where we have 
dropped the non logarithmic term which is
unimportant near the horizon.
(Note that $ \Delta_\lambda A= A-A'$ is always positive).
Upon postulating that the wave functions are WKB,
the Schwarzschild mode describing the shell is
 $\varphi_{\lambda,+} \simeq  e^{i S(A', A)}$ when the
gravitational interaction is taken into account

Our task is to now calculate the Bogoljubov transformation between
Unruh modes and Schwarzschild modes.
To impose the regularity of the Unruh modes at the horizon we 
 use the Eddington-Finkelstein coordinates
inside the shell 
$v = t +( {1 \over 2 \kappa'} \ln (r- r_{hor}) + r)$ and $r$, and then
impose analyticity in the upper complex $r$ plane.
In these coordinates eq. (\ref{Sl}) becomes
\begin{equation}
S_{shell}
 = -\Delta_\lambda A \kappa'  v / 8 \pi +  { \Delta_\lambda A 
\over 16 \pi} \ln (r -2 M') +
\int^{A}_{A'} {d \tilde A\over 16 \pi } \ln (r -  r_{hor}(\tilde A)) 
\end{equation}
up to non logarithmic terms. Analyticity in the lower half plane
imposes that the action on the right differs from the action on the
left by taking $r -2 M' \to (r - 2M')e^{ i \pi}$, 
hence $\ln (r- 2M') \to \ln (r- 2M') + i \pi$.
Since both the second and the third term 
equally contribute,
we obtain
\begin{equation}
S_{left} = S_{right} +i (\Delta_\lambda  A / 8)
\end{equation}
The Bogoljubov transformation between Unruh and Schwarzschild modes is
therefore
\begin{equation}
\phi_{\lambda} = N(\lambda) ( \varphi_{\lambda,+} + e^{-\Delta_\lambda   A / 8}
\varphi_{\lambda,-})
\label{phiB}\end{equation}
corresponding to the Bogoljubov coefficients
\begin{equation}
{|\beta_\lambda|^2 \over| \alpha_\lambda|^2 } = e^{- \Delta_\lambda   A/4}
\end{equation}
and one recovers eq. (\ref{R}).

Note that the two derivations lead to the same characterization
of the vacuum at the horizon.
There is indeed a
linear combination of 
{\it positive} frequency Unruh modes $\phi_{\lambda}$ eq. (\ref{phiB}) 
which at fixed $r$
has the $\Theta$ dependence of the 
form $De^{iC\omega e^{-\Theta}}$.

\section{Discussion}

We have shown that upon taking into account the gravitational back
reaction, the probability for a black hole to emit a quantum is
given by the exponential of the change in area of the black hole. The
appearance of this factor $\Delta A$ has the same origin as the
appearance of the term $d A$ in the first law of black hole mechanics,
namely the surface term $\Theta A/ 8 \pi$ at the horizon.
This derivation is not only valid for Schwarzschild black
holes. Indeed both derivations are also valid for charged black holes
and should be easily generalizable to rotating black holes. Furthermore
the first derivation which does not make appeal to any symmetry 
also applies to the transitions of 
uniformly accelerated detectors in Minkowski space
(the Unruh effect).

In obtaining this result we have made an assumption
about the physics at Planckian scales: we have postulated
that the specification of the vacuum state can be
implemented by the usual condition of regularity at the horizon 
when expressed in terms of the {\it local} 
coordinates $\Theta, U_{hor}$ and $r$.
In other words, we have assumed that the back-reaction
does not destroy the usual analytical characterization
of the vacuum.
One would hope to
have a justification for this assumption at a more fundamental level.
This probably requires a microscopic description of the physics at the
horizon at the Planckian scale, either 
string theory or some other quantum theory of gravity.

\vskip .3 truecm
{\bf Acnowledgements}
The authors would like to thank Roberto Balbinot and Ted Jacobson for 
enjoyable discussions.

\end{document}